# Sampling alien species inside and outside protected areas: does it matter?

Aristides Moustakas[1,*], Anneta Voutsela[2], and Stelios Katsanevakis[3]

1. Institute for Applied Data Analytics, Universiti Brunei Darussalam, Jalan Tungku Link, Gadong BE 1410, Brunei

2. School of Biological and Chemical Sciences, Queen Mary University of London, London, UK

3. Department of Marine Sciences, University of the Aegean, Mytilene, Greece

* Corresponding author

Aris Moustakas

Email: arismoustakas@gmail.com

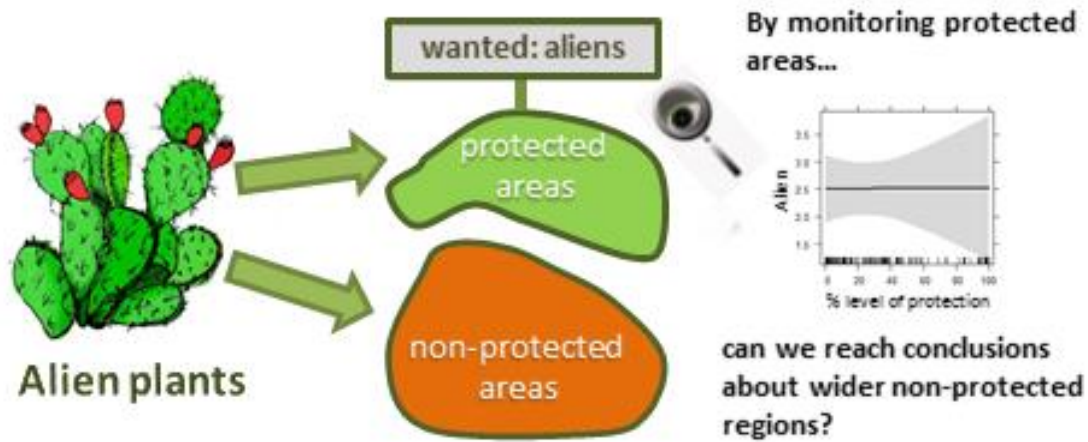

- We examine patterns of alien species richness inside and outside protected areas

- Alien species richness is lower in protected areas due to lower human activities

- Can we reach conclusions about non-protected regions by monitoring protected areas?

- Yes, we can, if the anthropogenic activity is accounted for

**Abstract**

Data of alien species presences are generally more readily available in protected than non-protected areas due to higher sampling efforts inside protected areas. Are the results and conclusions based on analyses of data collected in protected areas representative of wider non-protected regions? We address this question by analysing some recently published data of alien plants in Greece. Mixed effects models were used with alien species presences in 8.25 x 8.25 km cells as dependent variable and the percentage of protected area, as well as the agricultural and artificial land cover types richness (as indicators of human presence) as independent variables. In addition, the spatial cross-correlation between the percentage of protected area and alien species richness was examined across scales. Results indicated that the percentage of protected area per cell is a poor predictor of alien species richness. Spatial analysis indicated that cells with higher percentage of protected areas have slightly less alien species than cells with lower percentage of protected areas. This result is likely to be driven by the overall negative correlation between habitat protection and anthropogenic activities. Thus, the conclusions deduced by data deriving from protected areas are likely to hold true for patterns of alien species in non-protected areas when the human pressures are accounted for.

**Introductions**

Alien species are non-native taxa introduced by human agency to areas beyond their natural distribution and bio-geographical barriers [1]. Some alien species may become invasive, with important impacts on biodiversity, human health, and ecosystem services, through competition, predation, toxicity, transmission of pathogens, and the disruption of ecosystem functions [2-4]. With the human population being higher than ever before and increasing, together with unprecedented rates of mobility of humans and goods, the human assisted movement of living individuals or propagules beyond their natural distributions and across biogeographical barriers [5] has been accelerating [6]. Biological invasions are at the forefront of research in many disciplines such as ecology, conservation, epidemiology and food security [4, 7-9].

Protected areas are not immune to being invaded by alien species, and the risks can be high when this happens [10, 11]. Although management measures in protected areas and the expected increase of biodiversity and improvement of ecosystem functioning could control biological invasions, according to the 'biotic resistance' and 'diversity-stability' hypotheses [12], a number of studies has reported the opposite pattern, i.e. a positive correlation between alien and native species, according to the 'acceptance' hypothesis [10, 13].

To protect its habitats and species diversity, the European Union (EU) has created the Natura 2000 network of protected areas, which is one of the world's most extensive networks of conservation areas [14]. Alien species are generally better monitored in protected than non-protected areas due to existing monitoring frameworks and greater investment of monitoring and conservation efforts in such areas. Thus, rich datasets are more likely to be found in networks of protected areas than in unprotected areas - see e.g. [15-17].

A recent paper [16], investigated potential factors that influence alien plant species richness in the Natura 2000 sites in Greece. The main conclusions of that study were that native plant species richness and human population density have a positive effect on alien plant species presence. These findings are generally in agreement with a recent study examining hypotheses of vascular plant species invasibility in Crete and 49 surrounding islets [13]. The latter study examined alien species patterns in the Cretan area representing ~6.4% of the area of Greece (with ~24.5% of the Cretan area being protected), while [16] covered plant species throughout Greece but only in Natura areas, a ~16.3% of the area of Greece. How different would the patterns of alien plant species richness in Greece as recorded by [16] be outside protected areas? Are results and conclusions based on analyses of data collected in protected areas representative of the wider region? We herein investigate this question by analysing some recently published data of alien plants in Crete and surrounding islets as a case study.

**Methods**

*Dataset*

A dataset derived from digitising a plant atlas of the spatial distribution of native and alien vascular plant species from Crete and 49 surrounding islets (i.e. the Cretan area) was used [18, 19]. These data have been fully described in [13, 20] for addressing different questions. In brief the dataset was gridded in 162 square cells of 8.25 x 8.25 km. Coastal and inland cells have unequal land area; all inland cells have a total land area of 8.25 × 8.25 = 68.0625 km$^2$ while coastal cells and islets have smaller land areas (part of the cell is covered by the sea). In order to account for this effect, the total area of each cell was normalized by the area of inland cells. Each location (cell) in the study area was extensively surveyed for 10 years, up to three times each year, therefore the dataset is very reliable and species absences are likely to be real absences and not sampling artefacts [18, 19]. From the full plant dataset, alien species ($N_{alien}$ = 78) were defined [21].

The percentage of the area of each cell that is protected (i.e. included in the Natura 2000 network) was calculated, based on the distribution of protected areas [22]. Almost a fourth of the total Cretan area is within the Natura 2000, including all Cretan protected areas. Land cover data within each cell were classified using level three (the most detailed level) of the CORINE land cover classification system [22]. The land cover data of the Cretan area included 29 land cover types, of which 9 were agricultural, 7 were artificial, and 13 were natural. The agricultural, artificial, and natural land cover richness per cell was calculated as the number of these land cover types present on each cell respectively. Agricultural and artificial habitat types richness were used as indicators of human presences and pressures.

Climatic variables were derived from WorldClim [23] for the Cretan area. The original spatial resolution of the climatic data was 1 km with a temporal resolution of one month. The data were spatially re-scaled to 8.25 km in order to match with the grid size resolution of the plant atlas and temporally averaged per annum. The climatic variables used were: mean annual precipitation in mm year$^{-1}$ and mean annual temperature in °C per cell.

*Statistical analyses*

Linear mixed effects models [24] were used to investigate the relationship between alien species richness (dependent variable) and (i) the percentage of each cell in the Natura 2000 protected area network, (ii) the agricultural and (iii) the artificial land cover types richness within the cell, (iv) the mean annual precipitation, and (v) the mean annual temperature per cell (independent variables). The model also included the unique cell identity as a random effect to account for the fact that some coastal cells have unequal surface area than continental cells, and that there are also other underlying factors within each cell such as geographic and soil factors that are not accounted for in this analysis. In general, fixed effects account for the mean and random effects for the variance in the variables; see e.g. [25] for a similar rationale. Initially, agricultural and artificial habitat richness was compared with agricultural and artificial habitat cover as predictor variables of alien species richness; the former gave lower AIC values and was thereafter used in all analyses (see Supplement for details). Model selection was conducted using the AIC with maximum likelihood estimation [24]. Any deletion that did not increase AIC scores by more than 2 was deemed to be justified [26]. We ended up with the same model either by applying AIC or comparative F-tests. Inspection of residual plots for constancy of variance and heteroscedasticity indicated that the model was well behaved in all cases. Sequentially fixed effects were plotted with 95% confidence intervals. The analysis was performed using the 'nlme' and 'effects' packages in R [27].

We analysed the spatial cross-correlation between cells regarding alien species richness and the percentage of protected area using a multivariate spline cross-correlogram [28], in order to examine whether alien species richness increases with increased protected area. Spatial cross-correlogram estimates the spatial dependence at discrete distance classes. The region-wide similarity forms the reference line (the zero-line); the x-intercept is thus the distance at which objects are no more similar than that expected by-chance-alone across the region. This analysis examines in a spatially-explicit manner the correlation between two variables across scales [29]. On the vertical axis, cross-correlation values close to 1 indicate large positive correlation, while close to -1 large negative correlation, values close to zero indicate no correlation confidence intervals crossing zero are not significant, while on the horizontal axis distances of zero correspond to correlations within the same cell, distances of 1 correspond to correlations between a cell and the 8 neighbouring cells etc [29]. The cut-off range was set to 10 cells corresponding to a third of the maximum distance in the dataset; correlations beyond that distance (> 82.5 km) have little statistical and ecological meaning in particular as the study species are plants. Randomizations (N=999) were used to assess a 95% confidence interval. The analysis was performed using the 'ncf' package in R [27].

**Results**

Climatic variables (mean annual precipitation and mean annual temperature) were both not justified by AIC scores, and were not included in the most parsimonious model. The effect of the percentage of each cell in the Natura 2000 was both non-significant and with a slope close to zero (Fig 1a and Table 1); (non-significant, but kept into the final model for quantifying its effect). Agricultural and artificial land cover types richness (indicating human presence) were both significant (Table 1) and had a strong positive effect on alien species richness (Fig. 1b; 1c). When agricultural-type habitat richness increased by one (i.e. one more agricultural habitat type in the cell), there was a mean increase of 0.44 [Std.error = 0.18] in alien species richness (Table 1 and Fig. 1b), while when artificial-type habitat richness increase by one, there was a mean increase of 1.2 [Std.error = 1.99] in alien species richness (Table 1 and Fig. 1c). Overall alien species richness is not peaking at locations where the protected area is maximised (Fig 1d and 1e).

Spatial cross-correlation between the percentage of protected area on each cell and the alien species richness indicated a significant negative relationship at all distances below 10 cells (each cell corresponding at distances of 8.25 km) where the spatial cross-correlogram was cut off (Fig. 1f). The effect of the cross-correlation is relatively weak, correlation values are significant but equal to -0.12 at zero distances corresponding to comparisons between the same cell and increasing thereafter with values very close to zero at distances of 10 cells. Regression analysis indicated that there are less anthropogenic activities (lower agricultural and artificial land cover types richness, higher natural land cover types richness) in areas with high percentage of habitat protection (Fig. 2).

**Discussion**

Combining all results, (a) mixed effects model analysis, not accounting for space explicitly but including both the effects of habitat protection as well as the anthropogenic pressures indicates a negligible effect of habitat protection on alien species richness and a highly significant effect of anthropogenic pressures, (b) spatial analysis (accounting only for alien species richness and the percentage of protected area) indicates that alien species richness is lower inside than outside protected areas, and (c) there is a negative correlation between the percentage of protected area and anthropogenic pressures. Overall the percentage of protected area per cell is a poor predictor of alien species richness as deduced here using classical statistical analyses as well as in other studies with machine learning - see also Fig. 2 in [20].

Based on the data examined here, protected areas have slightly less alien species in cells with higher percentage of protected area than in cells with lower, in agreement with the findings of [30]. Combing the analysis performed here the percentage of protected habitat has a negligible effect on alien species (Fig. 1a), human presence has a strong positive effect on alien species richness (Fig. 1b & 1c), and alien species and percentage of protected area are negatively spatially correlated at finer scales (Fig. 1f).This result is likely to be driven by the overall negative correlation between habitat protection and anthropogenic activities (Fig. 2), as anthropogenic human activities inside protected areas [31] are lower than outside [32]. Thus the patterns deduced by [16] are likely to hold true for patterns of alien species in the rest of Greece (i.e. non-Natura areas) if the anthropogenic human pressures are accounted for, in the sense that it is the anthropogenic activity and not habitat protection *per se* driving the pattern. In protected areas the human pressure is lower, so the high human impacted sites are virtually missing in such areas even though such habitats may be common in European landscapes and belong to invaded habitats. This is not meant to imply that only data from protected areas should be used as the large and important part of variability (high disturbance part of the gradient) would not be covered, biasing the results.

In environmental sciences, ecology, and population biology it is common that datasets collected under different purposes than the current question of interest or under different experimental designs need to be merged together and analysed [33]. The statistical analysis of such data is often challenging due to missing values, sample sizes etc - see e.g. an application of capture-mark-recapture in plants in [34]. Conclusions deduced regarding biodiversity indices are often depended upon sampling effort and anthropogenic pressures [35, 36]. Unrecorded invasions are more common than anticipated [37]. The main conclusion of this work, the negligible effect of habitat protection on alien species richness, implies that datasets from protected areas can be used for deducing information regarding alien species presences on neighbouring non-protected areas when the human pressures have been accounted for in the analysis. This may often be a useful conclusion as in general protected areas are better monitored and sampled than non-protected areas, and thus more data are readily available. In addition, more data inside protected areas are likely to be generated in the future via citizen science [38] as well as with mining social media [39] with spatial information [40]. These data are more likely to derive from protected areas as these are areas of particular tourist interest and receive on average much higher number of visitors than non-protected areas.

The analysis and conclusions derived here regard the impact of habitat protection on alien species richness and show that this impact is negligible. Our results however, do not account for potential effects of alien species to native or endemic species or to the overall ecosystem functioning [41-44]. Our conclusion might differ in other regions with different levels of protection and different effectiveness of conservation measures. To assign causal positive correlation of protected areas on alien species, spatially and temporally replicated data are needed [45, 46] in order to deduce alien species trends since the establishment of protected areas. To that end dedicated databases (covering alien species in both protected and non-protected areas) and data mining are essential [47], as well as novel methods for their analytics [48].

**Acknowledgements**

We thank the artist Elena Sousta (Email: elenasousta@gmail.com) for designing the alien plant species image and artistic setup of the graphical abstract. Comments from three anonymous reviewers improved an earlier manuscript draft.

**Table 1.** Results from mixed effects models including alien species richness as dependent variable, and the percentage of protected area per cell, agricultural land cover richness per cell, and artificial land cover richness per cell as independent variables. To account for unequal surfaces of coastal cells, the unique cell's ID was included as a random effect.

| Variable | Value | Std.Error | DF | t-value | p-value |
|---|---|---|---|---|---|
| (Intercept) | -0.0217 | 0.6979 | 158 | -0.0311 | 0.9752 |
| Natura | 0.0001 | 0.0085 | 158 | 0.0162 | 0.9871 |
| Agric_Hab | 0.4411 | 0.1789 | 158 | 2.4655 | 0.0148 |
| Artif_Hab | 1.2037 | 0.1995 | 158 | 6.0338 | <0.001 |

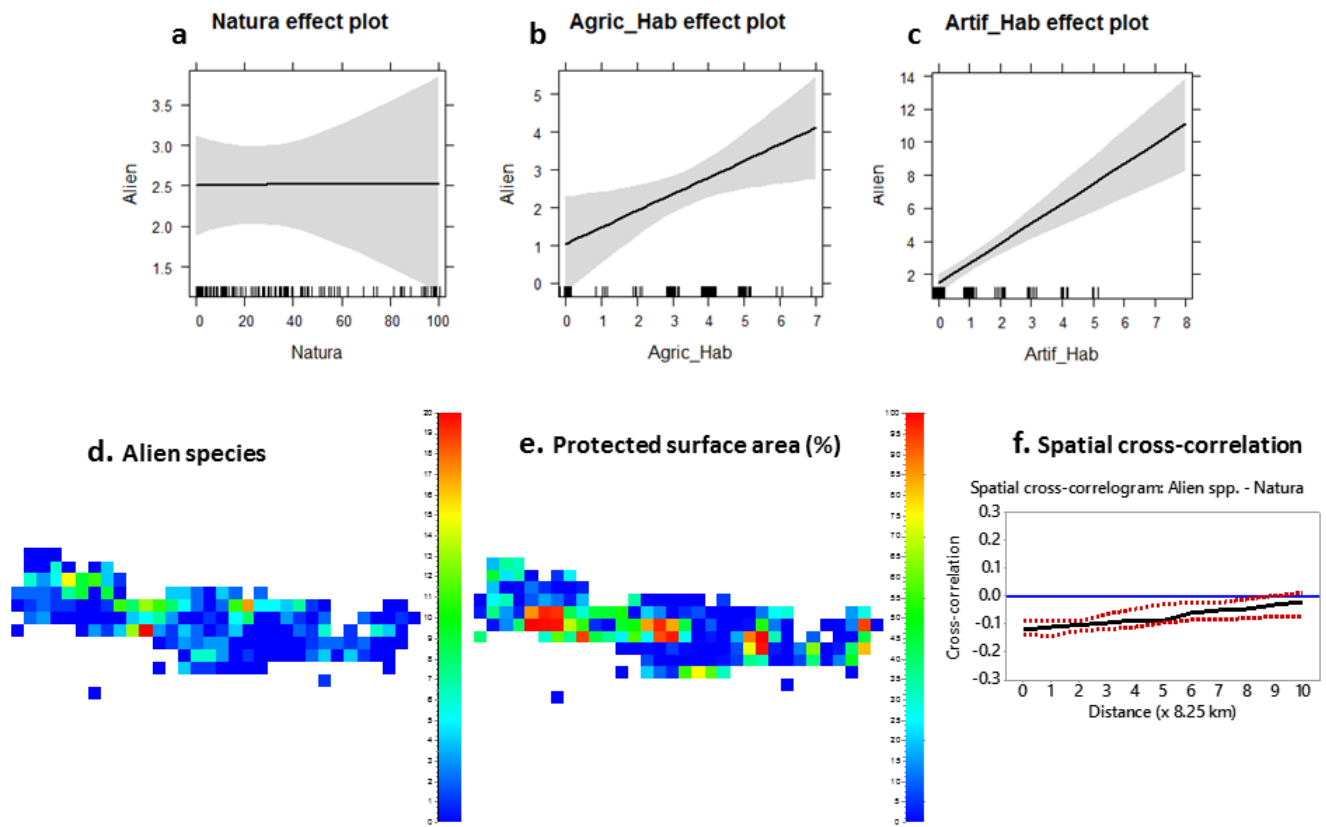

**Figure 1. (a - c)** Plotted effects of a mixed effects model with alien species richness as a dependent variable and the percentage of protected area on each cell (Natura %), the agricultural as well as the artificial land cover types richness per cell as independent variables. Grey shaded area indicates 95% confidence intervals. **(a)** The effects of Natura 2000 percentage on alien species richness. **(b)** The effects of agricultural land cover types richness on alien species richness. **(c)** The effects of artificial land cover types richness on alien species richness. **(d)** Spatial distribution of alien species richness on the Cretan area. Red shades of colours indicate higher richness values, while blue shades indicate lower. **(e)** Spatial distribution of the percentage of the Natura 2000 protected area on each cell. Red shades of colours indicate higher percentage of the cell's area within the Natura, while blue shades indicate lower**. (f)** Spatial cross-correlations between values of figures **(d)** and **(e)** across scales. The middle line indicates the fitted value while the upper and lower a 95% confidence interval. Each distance unit in the horizontal axis corresponds to one cell which is approximately equal to 8.25 km. The middle solid line is the fitted value while the dotted lines indicate the upper and the lower confidence interval respectively. If none of the three lines crosses zero, then the result is significantly positive (if all three lines are above zero) or significantly negative (if all three lines are below zero).

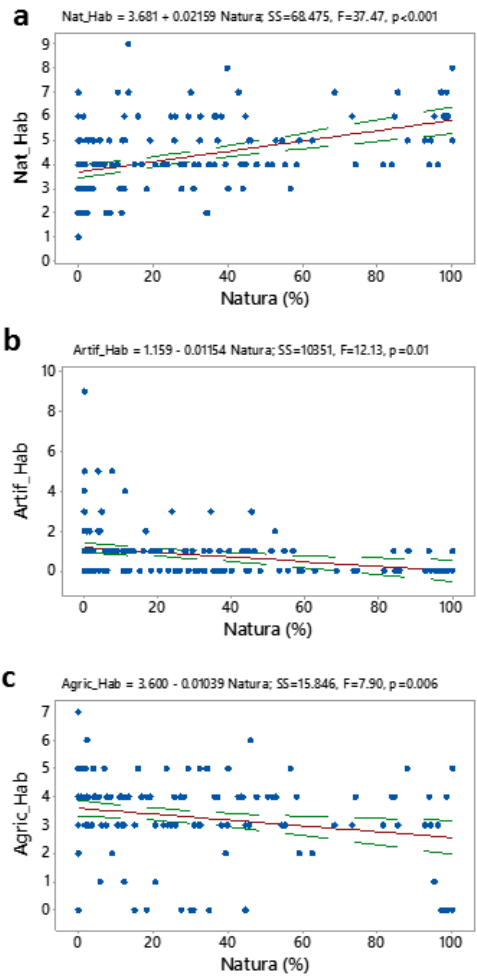

**Figure 2.** Linear regressions between the percentage of protected area (% of the cell within the *Natura* network), and **(a)** Natural, **(b)** Agricultural, and **(c)** Artificial land cover types richness.

# Supplementary material

**Details regarding mixed effects models building**

*Comparing agricultural and artificial land cover richness vs. agricultural and artificial land cover percentage of cover*

We initially tested whether alien species richness was best explained by artificial and agricultural land cover richness (model0.1) vs. artificial and agricultural land cover percentage of cover (model0.2) by choosing the model with the lowest AIC between richness or cover, all else been equal (with the climatic and Natura percentage of cover included as fixed effects and cell ID as a random effect in both models). All models were fitted using maximum likelihood estimation (ML) allowing comparisons between models with different fixed effects. The model that included artificial and agricultural land cover richness had the lowest AIC and therefore we proceeded with richness and not cover as predictor variables.

```
> model0.1<-lme(Alien~Natura+Agric_Hab+Artif_Hab+Temp_Mean+Precip_Mean,random =~1|Cell, data = alien2_data, method="ML")
> model0.2<-lme(Alien~Natura+Agric_Cover+Artif_Cover+Temp_Mean+Precip_Mean,random =~1|Cell, data = alien2_data, method="ML")
> anova(model0.1,model0.2)
         Model df      AIC      BIC    logLik
model0.1     1  8 831.6502 856.3510 -407.8251
model0.2     2  8 844.3031 869.0039 -414.1515
```

The fact that richness and not cover was the most parsimonious predictor of alien species richness in our view may be a result of analyzing alien species richness and not abundance as a dependent variable.

*Model selection – maximal model statistics*

The initial maximal mixed effects model structure included the five potential explanatory variables (percentage of Natura within the cell, agricultural land cover richness per cell, artificial land cover richness per cell, mean annual precipitation per cell, and mean annual temperature per cell), and cell ID as a random effect (independent variables) and alien species richness as dependent variable. The effects of both climatic variables (precipitation and temperature) were not justified by AIC scores and were removed from the most parsimonious model presented in Table 1 and Figure 1 in the main text. The effect of percentage of each cell within the Natura protected area network was kept in the final parsimonious model for quantifying its effect. The first section of the analysis below indicates an anova table of the full model with the last column indicating p-values. The summary table below the anova indicates model coefficients for each variable (Value) and its corresponding Standard error (Std.Error).

```
> model8<-lme(Alien~Natura+Agric_Hab+Artif_Hab+Temp_Mean+Precip_Mean,random =~1|Cell, data = alien2_data, method="ML")
> anova(model8)
            numDF denDF   F-value p-value
(Intercept)     1   156 109.43042  <.0001
Natura          1   156   5.28391  0.0229
Agric_Hab       1   156  18.16371  <.0001
Artif_Hab       1   156  37.18538  <.0001
Temp_Mean       1   156   2.17992  0.1418
Precip_Mean     1   156   3.19707  0.0757
>
> summary(model8)
Linear mixed-effects model fit by maximum likelihood
 Data: alien2_data
       AIC      BIC    logLik
  831.6502 856.351 -407.8251

Random effects:
 Formula: ~1 | Cell
        (Intercept) Residual
StdDev:    2.808673 1.053252

Fixed effects: Alien ~ Natura + Agric_Hab + Artif_Hab + Temp_Mean + Precip_Mean
                Value Std.Error  DF   t-value p-value
(Intercept) -6.064456 2.7658265 156 -2.192638  0.0298
Natura       0.003707 0.0098326 156  0.376983  0.7067
Agric_Hab    0.320013 0.1855556 156  1.724623  0.0866
Artif_Hab    1.229913 0.2054388 156  5.986765  0.0000
Temp_Mean    0.229409 0.1287317 156  1.782074  0.0767
Precip_Mean  0.003426 0.0019162 156  1.788036  0.0757
 Correlation:
            (Intr) Natura Agrc_H Artf_H Tmp_Mn
Natura      -0.361
Agric_Hab    0.043  0.128
Artif_Hab   -0.012  0.047 -0.287
Temp_Mean   -0.867  0.441 -0.150 -0.131
Precip_Mean -0.584 -0.199 -0.283  0.215  0.185

Standardized Within-Group Residuals:
       Min         Q1        Med         Q3        Max
-0.77069388 -0.21163888 -0.06044069  0.11984121  2.05583296

Number of Observations: 162
Number of Groups: 162
```